\documentclass[useAMS,usenatbib]{mn2e}
\usepackage{rotating}
\usepackage{graphics}
\usepackage{graphicx}
\usepackage{txfonts}
\usepackage{subfigure}
\usepackage{dblfloatfix}
\usepackage{fixltx2e}
\usepackage[usenames,dvipsnames]{xcolor}

\newcommand{\kms}{km\,s$^{-1}$}

\begin{document}

\title[Linear polarization in the cool active star II~Peg]
{Strong variable linear polarization in the cool active star II~Peg}

\author[L.~Ros\'{e}n et al.] 
{L.~Ros\'{e}n$^1$,
O.~Kochukhov$^1$, 
G.~A.~Wade$^2$ \\?
$^1$ Department of Physics and Astronomy, Uppsala University, Box 516, 751 20 Uppsala, Sweden \\
$^2$ Department of Physics, Royal Military College of Canada, PO Box 17000, Station Forces, Kingston, Ontario K7K 7B4, Canada}

\date{Accepted 2013 July 30. Received 2013 June 26; in original form 2013 June 03}

\pagerange{\pageref{firstpage}--\pageref{lastpage}}
\pubyear{2013}

\maketitle

\label{firstpage}

\begin{abstract}
Magnetic fields of cool active stars are currently studied polarimetrically using only circular polarization observations. Including linear polarization in the reconstruction of stellar magnetic fields allows more information about the magnetic field to be extracted and significantly improves the reliability of stellar magnetic field maps. The goal of this study is to initiate systematic observations of active stars in all four Stokes parameters and to identify cool stars for which linear polarization can be detected at a level sufficient for Zeeman Doppler imaging (ZDI). Four active RS CVn binaries, II~Peg, HR\,1099, IM~Peg, and $\sigma$~Gem, were observed with the ESPaDOnS spectropolarimeter at the Canada-France-Hawaii Telescope during a time period from February 2012 to January 2013. The least-squares deconvolution procedure was applied to derive mean polarization profiles of all four Stokes parameters. Linear polarization was detected in all four stars in at least one observation. At the same time, II~Peg showed an exceptionally strong and highly variable linear polarization signature throughout all observations. This establishes II~Peg as the first promising target for ZDI in all four Stokes parameters and suggests the feasibility of such an analysis with existing equipment for at least a few of the most active cool stars.
\end{abstract}


\section{Introduction}
\label{intro}

If a star possesses a magnetic field, the Zeeman effect leads to a broadening of the spectral lines in the integrated light (Stokes $I$ spectrum) and the appearance of signatures in the circular (Stokes $V$) and linear (Stokes $QU$) polarization spectra. If the field has a mean strength over the stellar surface of several kG, magnetically sensitive lines will be visibly split in slowly rotating stars \citep{Mathys1997}. But few stars possess this kind of field strength and therefore such magnetic splitting and broadening is difficult to identify. On the other hand, the circular polarization signatures are possible to detect even if the stellar magnetic field strength is significantly lower than one kG \citep{Donati97}, as typical for cool active stars. However, from basic theoretical considerations and observations of strong-field Ap stars \citep{Wade2000} Zeeman circular polarization is known to be about 10 times stronger than the associated linear polarization.

Most of the cool stars have relatively weak fields, making it difficult to detect any polarization, especially linear, in individual spectral lines. This problem can sometimes be overcome by using a multi-line technique, such as least-squares deconvolution (LSD) where the weak individual polarization signatures from thousands of spectral lines are combined into a single mean polarization profile \citep{Donati97}. This technique is very useful since it results in a manyfold increase of the S/N ratio of the mean profile despite of the assumptions that all combined lines are alike and scalable to one another.

If a star is observed at many rotational phases, the surface magnetic field topology can be reconstructed by applying the inverse modeling technique of Zeeman Doppler Imaging \citep[ZDI,][]{Brown91}. This analysis has been carried out many times, but always using only Stokes $V$ data even for the most active cool stars, such as II~Peg and HR\,1099 \citep{Donati1999,Petit2004,Kochukhov13}.

\begin{table*}
\caption{Log of spectropolarimetric observations of cool active stars in all four Stokes parameters. Uncertainties marked with * are mean values of the multiple measurements of that Stokes parameter.}
\label{tab1}
\centering
\resizebox{\textwidth}{!}{
\begin{tabular}{ccccccrcc}
\hline\hline
Star & Date  & HJD                 &   Stokes        & $t_{\rm exp}$ $\times$ 4 & $\sigma_{\rm LSD} \times 10^{-5}$ & \multicolumn{1}{c}{$\langle B_{\rm z} \rangle$} &  $\chi$$^2_\nu$ & FAP  \\
            & (UTC) & (2,400,000+) &$V$ $Q$ $U$     &   (s)                 &                    & \multicolumn{1}{c}{(G)}                   &               &           \\
\hline
II Peg   & 2012-07-07 & 56116.0331 &  1/1/1 & 825/825/825  &  2.5/1.8/1.8 &$104.9\pm2.4$   &258.6/10.1/32.4& $<$10$^{-16}$ / $<$10$^{-16}$ / $<$10$^{-16}$ \\
             
             & 2012-07-08 & 56117.0392 &  1/1/1 & 825/825/825  & 4.0/2.3/2.7 & $76.9\pm3.8$     &304.6/10.8/15.3& $<$10$^{-16}$ / $<$10$^{-16}$ / $<$10$^{-16}$ \\
             
             & 2012-07-09 & 56118.0285 & 1/1/1 & 825/825/825  & 3.6/2.1/2.4 & $63.3\pm3.3$     &478.5/11.8/23.3& $<$10$^{-16}$ / $<$10$^{-16}$ / $<$10$^{-16}$ \\

             & 2012-09-25 & 56195.7684 & 1/1/1 & 200/400/400  & 5.6/2.8/2.9 & $150.8\pm5.3$   &82.0/14.8/15.1& $<$10$^{-16}$ / $<$10$^{-16}$ / $<$10$^{-16}$ \\
             
             & 2012-09-26 & 56196.9182 & 1/1/1 & 200/400/400 & 5.0/2.8/2.8  & $51.5\pm4.8$    &100.0/11.0/27.0 & $<$10$^{-16}$ / $<$10$^{-16}$ / $<$10$^{-16}$ \\
             
             & 2012-09-27 & 56197.8587 & 1/1/1 & 200/400/400 &  5.0/2.7/2.6 & $84.0\pm4.7$     &280.3/7.4/29.3 & $<$10$^{-16}$ / $<$10$^{-16}$ / $<$10$^{-16}$ \\
      
             & 2012-09-28 & 56198.9808 & 1/1/1 & 200/400/400 &  5.9/3.4/3.1 & $55.3\pm5.5$      &79.5/9.3/19.7 & $<$10$^{-16}$ / $<$10$^{-16}$ / $<$10$^{-16}$ \\
   
             & 2012-09-29 & 56199.9629 & 1/2/1 & 200/400/400 & 6.2/3.9$^*$/3.3 & $-94.1\pm5.8$ &102.8/35.9$^*$/20.1 &$<$10$^{-16}$ / $<$10$^{-16}$ / $<$10$^{-16}$ \\
 
	      & 2012-09-30 & 56200.8659 & 1/1/1 & 200/400/400 & 7.1/3.6/3.8 & $-87.1\pm6.6$         &135.6/16.5/31.1  &$<$10$^{-16}$ / $<$10$^{-16}$ / $<$10$^{-16}$ \\

	      & 2012-10-01 & 56201.7619 & 1/2/1 & 200/400/400 & 7.2/3.4$^*$/3.4 & $90.1\pm6.8$   &144.1/8.3$^*$/26.7  &$<$10$^{-16}$ / $<$10$^{-16}$ / $<$10$^{-16}$ \\
	   
	      & 2012-12-31 & 56292.7619 & 1/1/1 & 200/400/400 & 5.5/3.0/2.9 & $10.9\pm5.2$            &88.3/10.1/28.2 & $<$10$^{-16}$ / $<$10$^{-16}$ / $<$10$^{-16}$ \\

             & 2013-01-02 & 56294.7334 & 1/3/1 & 200/400/400 & 6.6/4.9$^*$/3.5  & $-73.7\pm6.1$ &221.5/16.7$^*$/11.4 &$<$10$^{-16}$ / $<$10$^{-16}$ / $<$10$^{-16}$ \\

\hline
$\sigma$ Gem   & 2012-02-07 & 55964.8898 & 1/1/1 & 60/60/60 & 2.4/1.8/1.8 & $8.6\pm2.2$ &30.3/1.6/1.3& $<$10$^{-16}$ / $7.7\cdot10^{-3}$ / $6.0\cdot10^{-2}$   \\

                           & 2012-02-09 & 55967.0035 & 1/1/1 & 60/60/60 & 1.8/1.4/1.3 & $4.7\pm1.6$ &45.0/2.2/1.7 &$<$10$^{-16}$ / $6.6\cdot10^{-6}$ / $2.2\cdot10^{-3}$   \\

                           & 2012-02-11  & 55969.0619 & 1/1/1 & 60/60/60 & 2.7/2.0/1.9 & $-3.6\pm2.4$ &23.0/1.9/0.7 &$<$10$^{-16}$ / $1.8\cdot10^{-4}$ / $9.1\cdot10^{-1}$  \\

                           & 2012-02-13  & 55971.0232 & 1/1/1 & 60/60/60 & 2.4/1.7/1.8 & $-16.7\pm2.2$ &32.0/1.4/0.9 &$<$10$^{-16}$ / $4.4\cdot10^{-2}$ / $6.9\cdot10^{-1}$  \\

\hline
IM Peg & 2012-06-22  & 56101.1193 & 1/1/1   &210/210/210 & 2.5/1.8/2.0 & $50.9\pm2.6$    &50.4/2.5/4.1 &$<$10$^{-16}$ / $6.9\cdot10^{-8}$ / $<$10$^{-16}$                             \\

              & 2012-06-29  & 56108.1140 &1/1/1   &210/210/210 & 1.9/1.5/1.5 & $-11.6\pm2.0$ &59.4/1.3/1.2 &$<$10$^{-16}$ / $7.3\cdot10^{-2}$ / $1.4\cdot10^{-1}$   \\

              & 2012-07-09  & 56117.9495 &1/1/1   &210/210/210 & 2.3/2.1/2.2 & $-51.4\pm2.5$ &50.8/0.9/1.6 &$<$10$^{-16}$ / $6.3\cdot10^{-1}$ / $6.7\cdot10^{-3}$   \\

\hline
HR 1099 & 2012-07-09  & 56118.1047 & 1/1/1 & 200/200/200 & 3.3/2.4/2.3 & $-58.2\pm8.3$ &126.7/2.3/4.1 &$<$10$^{-16}$ / $4.9\cdot10^{-9}$ / $<$10$^{-16}$  \\
                
\hline
\end{tabular}}
\end{table*}
 
The Stokes $V$ parameter depends upon the magnetic vector projection onto the line-of-sight. This limits the information that can be extracted about the full magnetic field vector from circular polarization. For example, the azimuthal angle of the magnetic vector is undefined in the center of the stellar disk and the polar angle of the vector is undefined on the limb of the star. This may result in a complex interplay and cross-talks between radial, meridional and azimuthal components of the magnetic field depending upon its configuration and the phase coverage. As a result, ZDI based on Stokes $V$ spectra alone does not provide a fully reliable reconstruction of the stellar magnetic field with a complex topology. On the other hand, the Stokes $QU$ parameters depend upon the magnetic vector projected on the plane perpendicular to the line of sight, providing complementary information to Stokes $V$. Consequently, the quality of ZDI maps can be drastically improved if linear polarization is included in magnetic imaging \citep{Donati1997,Kochukhov02,Rosen12}.  

It is generally believed that, even with the aid of LSD, linear polarization is very difficult to detect for cool stars using existing spectropolarimeters. It has been argued that a high spectral resolution (above $10^5$) and more powerful line-averaging methods are needed for the analysis of linear polarization inside spectral lines \citep{Semel2009}. The study by \citet{Kochukhov11}, who reported detection of linear polarization for a couple of cool stars, did not disperse these concerns. Although their observations were made with the HARPSpol spectropolarimeter which provides a resolution of $R=109,000$, the resulting Stokes $Q$ and $U$ LSD profiles were characterized by a low signal-to-noise ratio, making them unsuitable for detailed modeling.

Here we present the first results of a pilot survey of linear polarization in a sample of cool active stars observed at a resolving power of $R=65,000$. Among our targets we found an object with a particularly strong linear polarization signal, potentially suitable for the cool-star ZDI in all four Stokes parameters.

\section{Observations}
\label{obs}
All the observations have been obtained with the ESPaDOnS spectropolarimeter \citep{Donati03} mounted at the Cassegrain focus of the Canada-France-Hawaii Telescope (CFHT). The thermally stabilized, bench mounted echelle spectrograph is fed by two optical fibers and has a spectral resolution of about $R = 65,000$ in its polarimetric mode and a wavelength coverage of 370--1050 nm for a single exposure. The polarimeter has three highly achromatic Fresnel rhombs of which one is a fixed quarter-wave rhomb and the other two are rotating half-wave rhombs. This set-up makes it possible to record both circular and linear polarization. A Wollaston prism is used to split the beam into two orthogonal polarization states, which are then dispersed by the spectrograph and recorded on the 2K$\times$4.5K E2V CCD detector. The polarized spectra were reduced automatically at CFHT by the Upena pipeline using the Libre-ESpRIT software \citep{Donati97}.

The observations were obtained on several separate occasions between February 2012 and January 2013. During the period from February to July 2012 we obtained Stokes $IQUV$ observations of all four stars. After preliminary analysis of these spectra II~Peg was established as the most interesting target and additional measurements of Stokes $IQUV$ were obtained for this star over the time span of three months. Each Stokes parameter observation consisted of a sequence of 4 sub-exposures with different orientations of the Fresnel rhombs \citep[e.g.][]{Donati97}.

Details about the individual ESPaDOnS Stokes $IQUV$ observations are summarized in Table~\ref{tab1}. The columns 1--5 give the name of the target, the observing date, a mean value of the Heliocentric Julian Date, the number of observations per Stokes parameter, and the sub-exposure times, respectively.

\section{Polarization profile analysis}
\label{an}
The four RS~CVn stars selected for this study are among the most magnetically active cool stars. According to previous studies, they show strong circular polarization in spectral lines. However, our photon noise limited polarization spectra would, although of relatively high S/N ratio, still not suffice to detect linear polarization signatures in individual spectral lines. To alleviate this problem we combined all available spectral lines into a single line profile by using the LSD technique \citep{Donati97}, which results in a significant increase of the S/N ratio. The LSD code developed by \citet*{Kochukhov2010} was used in this study.

LSD requires a line mask consisting of the wavelength, Land\'{e} factor, and central depth of each line. These parameters were extracted for lines in the wavelength range 400--890 nm from the {\sc vald} database \citep{Kupka1999} using {\sc marcs} model atmospheres \citep{Gustafsson2008} with the parameters used and determined in previous detailed studies of these stars (\citealt{Kochukhov13}; \citealt{Strassmeier2000}; \citealt*{Berdyugina99}; \citealt{Kovari01}). For all stars we used solar metallicity, except II~Peg which has [M/H]=\,$-0.25$. Effective temperature $T_{\rm eff}$=\,4750~K was adopted for II~Peg, HR\,1099, and $\sigma$~Gem, while $T_{\rm eff}$=\,4500~K was used for IM~Peg. The surface gravity of $\log g$=\,3.5 was used for II~Peg and HR\,1099 and $\log g$=\,2.5 for IM~Peg and $\sigma$~Gem. 

The final LSD masks used in our study contained 5500--8400 lines deeper than 20\% of the continuum. The resulting uncertainty per 2 \kms\ velocity bin of each LSD Stokes parameter is reported in column 6 of Table~\ref{tab1}. We calculated the mean longitudinal magnetic field $\langle B_{\rm z} \rangle$ \citep{Kochukhov2010} from the first moment of the Stokes $V$ profile and the results are listed in column 7 in Table~\ref{tab1}. 

In addition, we calculated the reduced chi-square $\chi^2_\nu$ and estimated the false alarm probability (FAP) \citep*{Donati1992} of the detection of polarization signatures in each Stokes parameter observation. We used $\nu=45$ degrees of freedom (velocity bins) for $\sigma$~Gem, $\nu=48$ for II~Peg and IM~Peg, and $\nu=66$ for HR\,1099. The $\chi^2_\nu$ values and FAP results are given in columns 8 and 9, respectively, in Table~\ref{tab1}.

\begin{figure}
\centering
\includegraphics[width=0.54\textwidth,angle=90]{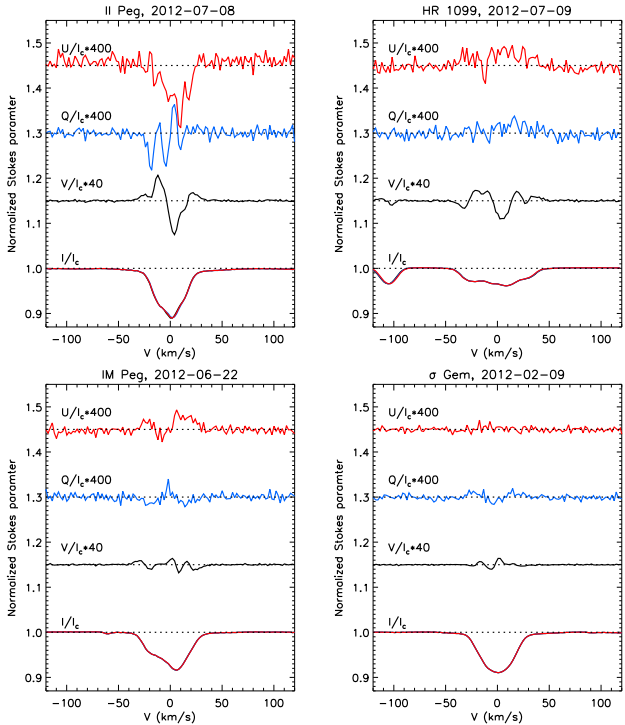}
\caption{Representative sets of the LSD Stokes $IQUV$ profiles for II~Peg, HR\,1099, IM~Peg, and $\sigma$~Gem. The polarization profiles are magnified and shifted vertically.}
\label{all_4s}
\end{figure}

In Fig.~\ref{all_4s} we show the mean polarization profiles with the lowest FAP. The Stokes $QUV$ spectra in Fig.~\ref{all_4s} are magnified by the same factors for all stars, allowing direct comparison between the four RS~CVn stars. It is evident that II~Peg shows by far the highest amplitude of the Stokes $QU$ signatures.

\section{Results}
\label{res}
\subsection{II~Peg}

\begin{figure*}
\centering
\includegraphics[scale=0.62]{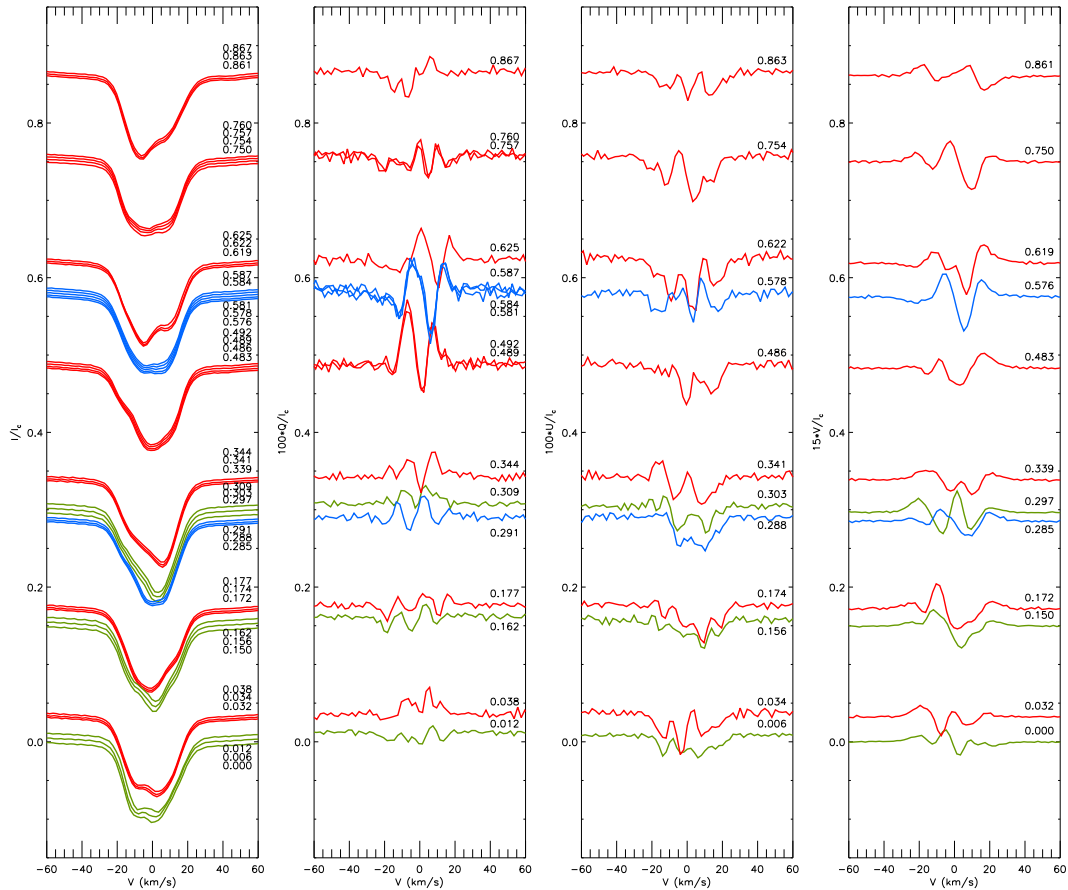}
\caption{LSD Stokes $IQUV$ profiles of II Peg. Each panel represents a different Stokes parameter. Profiles have been shifted vertically according to the rotational phase and the orbital radial velocity variation has been corrected for. Stokes $Q$ and $U$ have been magnified by a factor of 100 while Stokes $V$ has been magnified by a factor of 15. Different colors represent observations obtained in July 2012 (green), September/October 2012 (red), and December 2012/January 2013 (blue). 
}

\label{II-Peg}
\end{figure*}

This SB1 K2IV star is a well studied RS CVn star. Temperature maps have previously been produced \citep{Hackman2012} as well as magnetic maps from Stokes $V$ observations \citep{Kochukhov13}. In our study we found that this star exhibits an exceptional level of linear polarization in comparison to any other active cool star studied here or by \citet{Kochukhov11} for which linear polarization has been detected. All LSD polarization signatures obtained for II~Peg have a FAP below 10$^{-16}$. The phase variation of all LSD profiles is illustrated in Fig.~\ref{II-Peg}. The profiles chosen for Fig.~\ref{all_4s} correspond to phases 0.150--0.162 and their Stokes $QU$ amplitudes are actually among the lowest. 

Rotational phases of II~Peg were calculated using the orbital period of 6.724333~d \citep{Berdyugina98}, assuming tidal locking of the stellar rotation and binary motion. The overall phase coverage obtained in our study is good, allowing us to trace the evolution of magnetic spot signatures as the star rotates. However, these observations span over 6 months, making them unsuitable for ZDI due to surface structure evolution (e.g. compare the Stokes $V$ profiles at phases 0.285, 0.297 and 0.339).

The activity level of II~Peg appears to be continuously high throughout our observations. This star shows a long-term trend of magnetic activity. According to \citet{Kochukhov13}, the mean field intensity is currently increasing after passing through an activity minimum during 2008--2009.

All LSD Stokes $I$ profiles show distinct distortions due to temperature inhomogeneities. The shape of the Stokes $V$ profiles implies a complex field topology with large changes between phases and a significant intrinsic variation of the field structure over several months. The values of $\langle B_{\rm z} \rangle$ change sign from one phase to the next and vary in absolute value between about 11~G and 150~G with a typical 1 $\sigma$ uncertainty of about 5~G. The amplitude of Stokes $V$ is typically $\sim$\,$2\cdot10^{-3}$~$I_{\rm{c}}$. The LSD Stokes $QU$ profiles are also highly variable in shape and amplitude. The typical amplitude is $\sim$\,$4\cdot10^{-4}$~$I_{\rm{c}}$ while the largest reach as high as $8\cdot10^{-4}$~$I_{\rm{c}}$ for Stokes $Q$ at phases 0.5--0.6. This implies that the Stokes $QU$ amplitude is 5--10 times smaller than that of Stokes $V$ in LSD profiles.

\subsection{HR\,1099}
HR\,1099 is another very active RS~CVn star. Previous Doppler imaging studies found spot structures with temperatures more than 1000~K below the photospheric value \citep{Strassmeier2000}. Its magnetic field has also previously been mapped using Stokes $V$ only \citep{Donati1999,Donati2003,Petit2004}. Linear polarization has been detected for HR\,1099 by \citet{Kochukhov11} using $R>10^5$ observations with the HARPSpol spectropolarimeter. The single observation of HR\,1099 obtained here shows a similarly convincing detection of linear polarization in spectral lines, but this signal is far less prominent as compared to II~Peg. 

The Stokes $I$ profiles of HR\,1099 are distorted by the temperature spots. The Stokes $I$ signature of the secondary star is also clearly visible. The Stokes $V$ profile is relatively strong and complex with $\langle B_{\rm z} \rangle$\,=\,$-58.2\pm8.3$~G. A weaker Stokes $V$ signal from the secondary can also be seen. The Stokes $QU$ profiles both show a definitely detected signature within the mean line of the primary with an amplitude of about $10^{-4}$~$I_{\rm{c}}$, which is 10 times weaker than Stokes $V$. 

\subsection{IM~Peg}
The single-lined RS CVn star IM~Peg was chosen to be the guide star for the Gravity Probe B mission and hence has been studied meticulously {\citep[e.g.][]{Zellem2010}}. It is known to have cool temperature spots \citep{Berdyugina2000} although its magnetic field has not been mapped.

The rotational period of the primary and the orbital period are both close to 24.5~d \citep{Fekel99,Marsden05}. Our three observations of this star were obtained over 18~d and hence correspond to significantly different rotational phases.

The LSD Stokes $I$ profiles of IM~Peg again show signs of temperature inhomogeneities. A weak contribution of the secondary star can be seen on the blue side of the primary profile illustrated in Fig.~\ref{all_4s}. Circular polarization of a similar amplitude, $\sim$\,$4\cdot10^{-4}$~$I_{\rm{c}}$, was detected in all three observations with $\langle B_{\rm z} \rangle$ ranging between about $-50$~G to $+50$~G with a typical 1 $\sigma$ uncertainty of just above 2~G. The Stokes $QU$ profiles in Fig.~\ref{all_4s} are clearly visible and the detection level is as secure as for HR\,1099. The amplitude of this signal is $\sim$\,$10^{-4}$~$I_{\rm{c}}$, implying that for this rotational phase the linear polarization was 4 times weaker than the circular polarization. However, the other two observations of IM~Peg do not show as prominent linear polarization signatures despite a similar quality of observations.  

\subsection{$\sigma$~Gem} 
This RS~CVn star has enhanced chromospheric activity and surface temperature spots \citep{Hatzes93,Kovari01}. The rotational period is about 19.6 days \citep*{Strassmeier1999} and since our four observations were obtained over 6 nights in February 2012, they all represent different rotational phases. 

Compared to other stars, the Stokes $I$ LSD profiles of $\sigma$~Gem does not show major distortions due to temperature spots. The signal in all Stokes $V$ profiles is detected with high certainty and has a fairly complex shape with a typical amplitude of $\sim$\,$4\cdot10^{-4}$~$I_{\rm{c}}$. The value of $\langle B_{\rm z} \rangle$ varies between $-17$~G and +9~G with a typical 1 $\sigma$ uncertainty of about 2~G. Only a single observation in Stokes $Q$ showed high significance of the signal with the FAP\,$<$\,10$^{-5}$. The amplitude of this signature is about $6\cdot10^{-5}$~$I_{\rm{c}}$, i.e. somewhat more than 10\% of the Stokes $V$ amplitude. On the other hand, the Stokes $U$ signal was not securely detected in any of our observations.

\section{Discussion}
\label{dis}

Modeling of cool-star magnetic field topologies has so far been limited to interpretation of unpolarized and circular polarization spectra. Detecting linear polarization in such stars was believed to be very difficult or impossible with the current spectropolarimeters at 4-m class telescopes. A recent study by \citet{Kochukhov11} has succeeded in detecting weak $Q$ and $U$ signatures in HR\,1099 and $\varepsilon$~Eri using observations at a very high resolution, $R > 10^5$. However, these detections had a S/N ratio insufficient for ZDI. 

Our study significantly increases the sample of cool stars observed in all four Stokes parameters. We demonstrate that it is indeed possible to detect linear polarization in cool active stars at a level suitable for magnetic field reconstruction. We also show that this can be done at a spectral resolution significantly lower than $10^5$, ($R=65,000$ in our study). Observations of this quality allowed us to detect Stokes $QU$ signal at least once for each of the four RS~CVn stars considered in our study. Furthermore, the linear polarization was detected for all observations of II~Peg at the level of $>10\sigma$. The amplitude of the Stokes $QU$ signal in this star is as large as the amplitude of circular polarization detected in IM~Peg and $\sigma$~Gem. Thus, II~Peg is established as a very promising candidate for magnetic Doppler imaging in all four Stokes parameters.

Even though II~Peg was known to be one of the most magnetically active stars, it was not expected to show such a strong linear polarization. Previous circular polarization studies {\citep[e.g.][]{Donati1992,Donati97}} demonstrate that the Stokes $V$ signatures of II~Peg are only marginally stronger than those of similar RS~CVn stars, e.g. HR\,1099, and the surface field strength reconstructed with ZDI is comparable \citep{Petit2004,Kochukhov13}. Despite this, we found that the Stokes $Q$ and $U$ profiles of this star are up to 13 times stronger than for the other three targets. 

The activity level of II~Peg was high during all 6 months when we observed it. This star showed particularly strong Stokes $Q$ signatures at phases 0.5--0.6. The observations corresponding to these phases were obtained 3 months apart, but they retained a similar strength and shape. This characteristic, combined with the overall consistent detection and coherent morphology and variation of the linear polarization profiles, convincingly supports that this signal is due to stellar magnetic field rather than an instrumental artifact. In general, the ESPaDOnS spectropolarimeter is especially well-suited for four Stokes parameter observations of cool stars since the cross-talks between Stokes parameters are regularly monitored and, after several upgrades, are maintained at the level below 1\% \citep{Silvester2012}.

The common notion that linear polarization is about 10 times weaker than circular polarization might be somewhat pessimistic. The linear polarization detected in II~Peg and IM~Peg is stronger than that. Furthermore, the Stokes $V$ profile of IM~Peg is smaller compared to that of HR\,1099 shown in Fig.~\ref{all_4s}, but their respective linear polarization amplitudes are still similar. The 10\% estimate is still a reasonable guideline when trying to detect linear polarization in cool active stars.

The methodology of interpreting the mean Stokes $Q$ and $U$ signatures remains to be developed. Our study shows that, similar to the four Stokes parameter observations of Ap stars, the LSD method is capable of extracting high-quality mean profiles for magnetic stars with much more complex fields. Therefore, there is no reason to abandon LSD in the context of four Stokes parameter studies. However, the corresponding modeling cannot be carried out under the usual assumption of circular polarization studies because the LSD Stokes $Q$ and $U$ spectra do not behave as a spectral line with average parameters \citep{Kochukhov2010}. In general, the Stokes $Q$ and $U$ signatures are less alike for different spectral lines and cannot be easily rescaled to the mean LSD profile in comparison to Stokes $V$. A new methodology, probably based on the full polarized radiative transfer calculations covering all lines included in the LSD mask, needs to be developed and tested. This will be the focus of future research.

\section*{Acknowledgements}
OK is a Royal Swedish Academy of Sciences Research Fellow, supported by the grants from Knut and Alice Wallenberg Foundation and Swedish Research Council. GAW is supported by a Discovery Grant from the Natural Science and Engineering Research Council of Canada (NSERC).

\bibliographystyle{mn2e}
\bibliography{astro_ref_v1}

\label{lastpage}

\end{document}